\documentclass{emulateapj}
\usepackage{amsmath}
\newcommand{\be}{\begin{equation}}
\newcommand{\ee}{\end{equation}}

\newcommand{\der}{{\rm d}}
\newcommand{\erf}{{\rm erf}}
\newcommand{\m}{_{\rm m}}
\newcommand{\n}{_{\rm n}}
\newcommand{\xm}{x_{\rm m}}
\newcommand{\cc}{_{\rm c}}
\newcommand{\cz}{_{\rm c0}}
\newcommand{\s}{_{\rm s}}

\newcommand{\io}{_{\rm a}}
\newcommand{\ii}{_{\rm i}}

\newcommand{\tc}{\tau_{\rm cr}}

\newcommand{\rs}{r_{\rm s}}
\newcommand{\mr}{{\cal R}_{\rm m}}
\newcommand{\ra}{{\cal R}_{\rm a}}

\newcommand{\nbody}{$N$-body\ }
\newcommand{\modotm}{~M_\odot}
\newcommand{\delm}{\Delta_{\rm m}}
\newcommand{\delv}{\Delta_{\rm vir}}

{

}

\shorttitle{DENSITY PROFILE AND CENTRAL BEHAVIOR OF HALOS}
\shortauthors{SALVADOR-SOL\'E ET AL.}
\slugcomment{The Astrophysycal Journal, 647:000-000, 2007 September 20}

\begin{document}

\title{The nature of dark matter and the density profile and central
behavior of relaxed halos}

\author{Eduard Salvador-Sol\'e$^{1}$, Alberto Manrique$^{1}$,\\
Guillermo Gonz\'alez-Casado$^2$, and Steen H. Hansen$^3$}

\affil{$^1$ Departament d'Astronomia i Meteorologia, and
Institut de Ci\`encies del Cosmos (UB/IEEC) \\ associated with the
Instituto de Ciencias del Espacio (CSIC), Universitat de Barcelona, Spain}

\affil{$^2$  Departament de Matem\`atica Aplicada II, Universitat Polt\`ecnica de Catalunya, Spain}

\affil{$^3$ Dark Cosmology Centre, Niels Bohr Institute, University of  Copenhagen, Denmark}



\begin{abstract}

We show that the two basic assumptions of the model recently proposed
by Manrique and coworkers for the universal density profile of cold
dark matter (CDM) halos, namely that these objects grow inside out in
periods of smooth accretion and that their mass profile and its radial
derivatives are all continuous functions, are both well understood in
terms of the very nature of CDM. Those two assumptions allow one to
derive the typical density profile of halos of a given mass from the
accretion rate characteristic of the particular cosmology. This
profile was shown by Manrique and coworkers to recover the results of
numerical simulations. In the present paper, we investigate its
behavior beyond the ranges covered by present-day \nbody
simulations. We find that the central asymptotic logarithmic slope
depends crucially on the shape of the power spectrum of density
perturbations: it is equal to a constant negative value for power-law
spectra and has central cores for the standard CDM power spectrum. The
predicted density profile in the CDM case is well fitted by the 3D
S\'ersic profile over at least 10 decades in halo mass. The values of
the S\'ersic parameters depend on the mass of the structure
considered. A practical procedure is provided that allows one to infer
the typical values of the best NFW or S\'ersic fitting law parameters
for halos of any mass and redshift in any given standard CDM
cosmology.
\end{abstract}




\section{INTRODUCTION}\label{intro}

The universal shape of the spherically averaged density profile
of relaxed dark halos in high-resolution \nbody simulations is
considered one of the major predictions of standard cold dark matter
(CDM) cosmologies. Down to 1 \% of the virial radius $R$, it is
well fitted by the so-called NFW profile \citep{NFW96,NFW97},
\begin{equation}
\rho(r)=\frac{\rho\cc\rs^3}{r(\rs+r)^2}\,,
\label{NFW}
\end{equation}
specified by only one mass-dependent parameter, the scale
radius $\rs$ or equivalently the concentration $c\equiv \rs/R$.
 
At radii smaller than 1 \% of the virial radius, however, the behavior
of the density profile is unknown. On the basis of recent numerical
simulations, some authors advocate a central asymptotic slope
significantly steeper (Moore et al. 1998; Jing \& Suto 2000) or
shallower \citep{TN01,R03,HS06} than that of the NFW law. Others
suggest an ever decreasing absolute value of the logarithmic slope
\citep{P03,N04,R05}, which might tend to zero as a power of the radius
as in the three-dimensional (3D) S\'ersic (1968) or Einasto (Einasto
\& Haud 1989) laws \citep{M05,M06}.

This uncertainty is the consequence of the fact that the origin of such
a universal profile is poorly understood.  Two extreme points of view
have been envisaged. In one of these, it would be caused by repeated
significant mergers \citep{sw98,rgs98,SCO00,DDH03}, while in the other
it would be essentially the result of smooth accretion or secondary
infall \citep{arfh,ns99,pgrs,kull99,M03,wbd04,As04}.

The fact that the $M-c\/$ relation at $z=0$ is consistent
(Salvador-Sol\'e et al.~1998,
\citealt{Wechs02,Zetal03a}) with the idea that all halos emerge from
major mergers with similar values of $c$, which then decreases
according to the inside-out growth of halos during the subsequent
accretion phase, seems to favor an important role of mergers. However, the
purely accretion-driven scenario is at least as attractive, as the
inside-out growth during accretion leads to a typical density profile
that appears roughly to have the NFW shape with the correct $M-c\/$
relation in any epoch and cosmology analyzed (Manrique et al.~2003,
hereafter M03).

Certainly the effects of major mergers cannot be neglected in
hierarchical cosmologies, so both major mergers and accretion should
contribute in shaping relaxed halos. However, as noted by M03, if
the density profile arising from a major merger were set by the
boundary conditions imposed by {\it current} accretion, then the
density profile of halos would appear to be independent of their past
aggregation history, so halos could be assumed to grow by pure
accretion without any loss of generality. All the correlations shown
by relaxed halos in numerical simulations can be recovered under this
point of view \citep{SMS}. Simultaneously, this would explain why
halos with very different initial conditions and aggregation histories
have similar density profiles (\citealt{RD06}).

In the present paper, we show that the M03 model relies on two basic
assumptions, namely, (1) that halos grow inside out in periods of smooth
accretion, and (2) that the mass profile and all its derivatives are
continuous functions. The former assumption is supported by the
results of numerical simulations (\citealt{SMS,L06,RD06,RD07}), while
the second one is at least not in contradiction with them. In the
present paper we show that both assumptions are in fact sound from a
theoretical point of view, as they can be related to the very nature of
CDM. This renders the predictions of the model beyond the range of
current simulations worth examining in detail, as is done below.

For simplicity, we consider spherical structures, which at best
is an approximation to the triaxial structures observed in numerical
simulations. Secondary infall is known to be influenced by deviations
from spherical symmetry \citep{BM96,ZNH96}. Yet, in an accompanying
paper \citep{GC07}, we show that this does not seem to affect the
fundamental role of the two assumptions given above. Another
simplifying assumption used here is the neglect of substructure. It
has recently been shown in high-resolution simulations that about 57
\% of the halo mass is collected in the previous major merger
\citep{Fea05}. However, substructures (and sub-substructures) contribute
only about 5 \% of the total mass in halos \citep{DKM07} whose density
profile is well described by equation (\ref{NFW}). It seems therefore a
good first approximation to ignore substructure.

The paper is organized as follows. In \S \ref{rev}, we show how the
M03 model emerges from the properties of standard CDM. The
behavior of the predicted density profile at extremely small radii and
for a wide range of halo masses is investigated in \S \ref{dens}. Our
results are summarized in \S \ref{sum}.

\section{CDM properties and halo density profile}\label{rev}

\subsection{Inside-out growth during accretion}\label{relform}

We now argue that halos grow inside out during accretion, as is indeed
found in numerical simulations, because of some properties of CDM; in
particular, its characteristic power spectrum, leading to a slow halo
accretion rate. As shown in \S \ref{boundary}, this has important
consequences for the inner structure of these objects.

Schematically, one can distinguish between minor and major mergers. In
minor mergers, the relative mass increase produced, $\Delta\equiv
\Delta M/M$, is so small that the system is left essentially
unaltered, whereas in major mergers $\Delta$ is large enough to cause
rearrangements. 

\begin{figure}[htb]
	\centering
	\includegraphics[angle=0,width=0.5\textwidth]{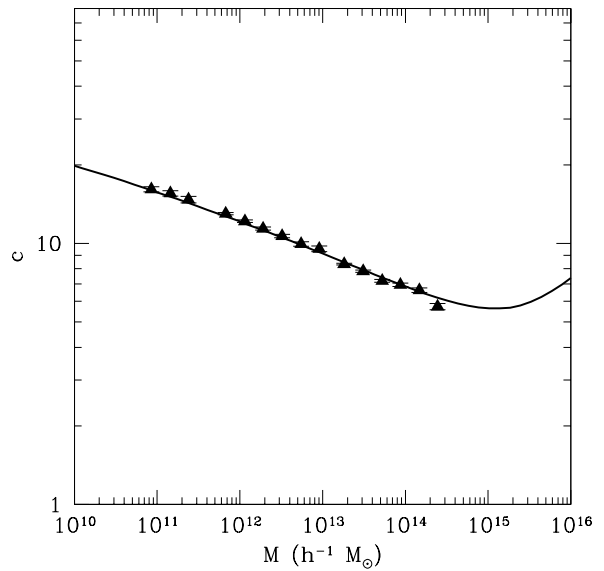}
	\vspace{-1 cm}
	\caption{$M-c\/$ relation at $z=0$ predicted in the concordance model 
by the M03 model for $\delm=0.26$ ({\it solid line}) compared to the
empirical relation traced by those points, with minimal error bars
obtained by \citet{Zetal03b} from high-resolution simulations
({\it triangles with error bars}).}
	\label{fit}
\end{figure}

The smaller the value of $\Delta$, the most frequent the mergers \citep{LC93}. For
this reason, although individual minor mergers do not affect the
structure of halos, their added contribution yields a ${\it smooth}$
secular mass increase, the so-called accretion, with apparent effects
on the aggregation track $M(t)$ of the halo. The accretion-scaled
rate, $\dot M/M(t)$, is given by (\citealt{rgs01}, hereafter RGS01)
\begin{equation}
\ra(M,t) = \int_0^{\delm} \der \Delta\; \Delta\; 
\mr(M,t,\Delta)\,,
\label{accrate} 
\end{equation}
where $\mr(M,t,\Delta)$ is the usual Lacey-Cole \citep{LC93} instantaneous
merger rate (see eq.~[\ref{mrate}]) and $\delm$ is the maximum value
of $\Delta$ for mergers contributing to accretion. In contrast, 
less frequent major mergers yield
notable sudden mass increases or {\it discontinuities} in $M(t)$.

After undergoing a major merger (and virializing), halos evolve as
relaxed systems until the next major merger. The fact that standard
CDM is {\it nondecaying and non-self-annihilating}\footnote{Both
decay and annihilation rates are extremely small for realistic dark
matter particle candidates.}  guarantees that the mass collected
during such periods is conserved. This does not yet imply that halos
grow inside out during accretion, because their mass distribution
might still vary due to energy gains or losses or to the action of
accretion itself. Even if each individual minor merger would leave the
halo unchanged their collective action might alter these systems.
However, standard CDM is also {\it dissipationless} and therefore
halos cannot loose energy. Furthermore, under the assumption of
spherical symmetry halos cannot suffer tidal torques from surrounding
matter, and hence the surrounding matter is unable to alter the
kinetic energy of the halo. Therefore, the only possibility for a
time-varying inner mass distribution of accreting halos is that the
accretion process causes it itself.

\begin{figure*}[htb]
	\centering
	\includegraphics[angle=0,width=0.8\textwidth]{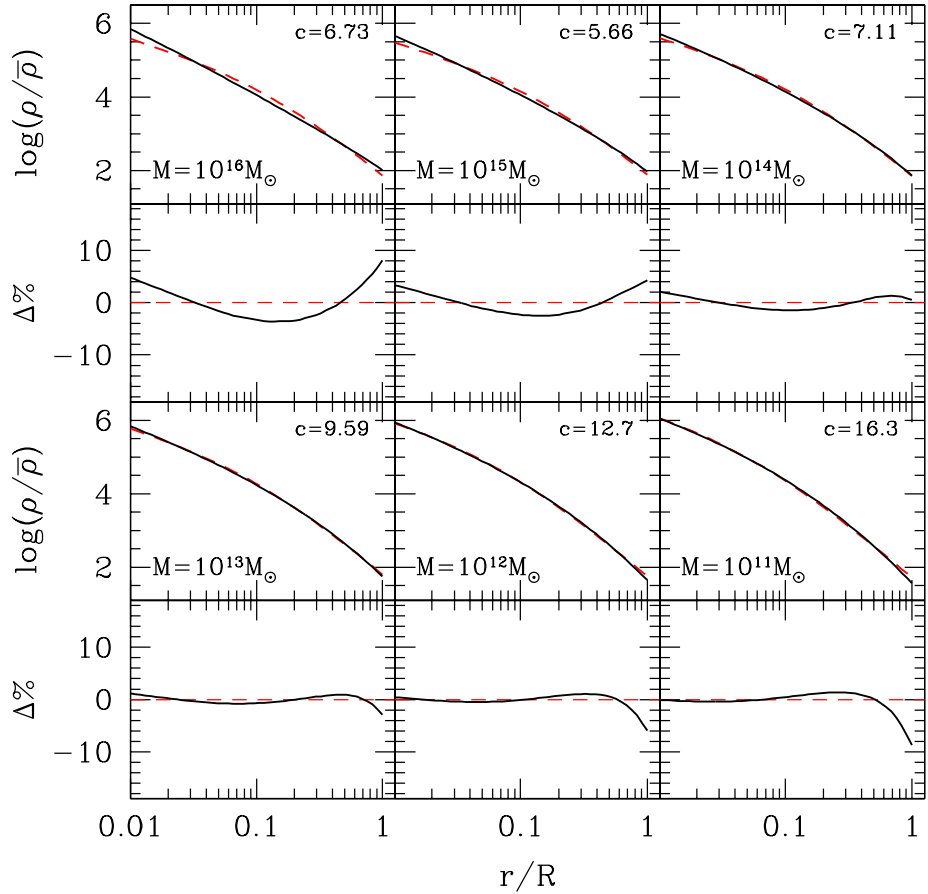}
	\vspace{-1.2cm}
	\caption{Predicted density profiles ({\it solid lines}) compared to
their fits to a NFW profile ({\it dashed lines}) for halo masses at $z=0$,
ranging from $10^{16}\modotm$ to $10^{11}\modotm$ (or, equivalently,
from $10^3$ to $10^{-2}$ times the current critical mass $M_{\star0}$
for collapse), in the concordance model. {\it Upper subpanels}:
Profiles. {\it Lower subpanels}: Residuals of the logarithmic fits. The
halo mass and the corresponding best-fitting value of the NFW
concentration parameter are quoted in each panel.}
	\label{nfw}
\end{figure*}

This possibility, that the process of accretion itself could alter the
internal mass distribution, would be realized only if the accretion
time $1/\ra$ were smaller than the dynamical time $\tc$. On the
contrary, if $1/\ra$ is substantially larger than $\tc$, the adiabatic
invariance of the inner halo structure will be guaranteed, and the
halo will evolve inside out. Thus, by requiring $1/\ra$ to be $C$
times larger than $\tc$, we are led to the equation
\begin{equation}
C\,\left[\frac{4\pi}{3} G\delv(t)\bar\rho(t)\right]^{1/2}=
\ra(M,t)\,,
\label{cond}
\end{equation}
which gives the upper mass $M\io$ for inside-out growth at $t$. In
equation (\ref{cond}), $\bar\rho$ is the mean cosmic density and
$\delv(t)$ is the virialization density contrast given e.g. by
\citet{bn98}. Here $M\io$ is indeed an upper limit because $\ra(M,t)$ is
an increasing function of $t$; see RGS01. In any CDM cosmology analyzed,
equation (\ref{cond}) appears to have no solution for values of $C$
significantly larger than unity in the relevant redshift range. This
means that accretion is always slow enough for halos to grow
inside out, as required by the M03 model. 

As previously mentioned, the inside-out growth of halos in accreting
periods is unambiguously confirmed by the results of \nbody
simulations (\citealt{SMS,L06,RD06,RD07}). It is also consistent with
the fact that dark matter structures preserve the memory of initial
conditions, in the sense that the most initially overdense regions end
up being the central regions of the final structures \citep{DMM05},
implying that the spatial positions of particles are not significantly
perturbed by merging/accretion during the assembly of the
structures. Likewise, the energy of the individual particles in the
final structure (at $z=0$) is very strongly correlated with their
energies at much earlier times ($z=10$; \citealt{DR06}). This shows
that particles even preserve the memory of the initial energies.

\subsection{Smoothness of the mass profile}\label{boundary}

Contrarily to an ordinary fluid, CDM is {\it collisionless\/} and {\it
free-streaming} and, hence, cannot support discontinuities (shock fronts) in
the spatial distribution of any of its macroscopic properties.  As a
consequence, all radial profiles in relaxed halos are necessarily
smooth. This holds in particular for the mass profile, $M(r)$, and its
radial derivatives,\footnote{We are naturally referring to the {\it
theoretical} mass profile, equal to the average profile over random
realizations or over time realizations as long as the halo evolves
inside-out.} which has the following consequence.

The inside-out growth of a halo during accretion (\S \ref{relform})
implies that the mass profile $M(r)$ built at that interval is the
simple conversion, through the definition of the instantaneous virial
radius
\begin{equation}
R(t)= \left[\frac{3M(t)}{4\pi \delv(t)\bar \rho(t)}\right]^{1/3}\,,
\label{r}
\end{equation}
of the associated mass aggregation track $M(t)$. 

The smoothness condition implies that the old $M(r)$ profile must
match perfectly with the new part of the profile built during that
time. Since minor mergers only cause tiny discontinuities, the new
piece of the $M(t)$ track that they produce is well approximated by a
smooth function, and, as the functions $\bar \rho(t)$ and $\delv(t)$
in equation (\ref{r}) are also smooth functions, the corresponding
piece of the $M(r)$ profile automatically fulfills the right
smoothness condition.  Thus, the system can grow, during accretion,
without the need to essentially rearrange its structure.

Only when a halo undergoes a major merger and its $M(t)$ track suffers
a marked discontinuity will the mass profile prior to the major merger
no longer match the piece that begins to develop after it. Since
the $M(r)$ profile cannot have any discontinuities, the halo is then
forced to rearrange its mass distribution (through violent relaxation)
to fulfill the required smooth condition.

In other words, the fundamental assumption of the M03 model that the
mass distribution of relaxed halos is determined by their current
accretion rate (through dramatic rearrangements of the structure on the
occasion of major mergers and very tiny and negligible ones during
accretion periods), is simply the natural consequence of the slowly
accreting, nondecaying, nonself-annihilating, dissipationless, and
collisionless nature of standard CDM.

\subsection{The M03 model}\label{M03}

The mass profile of a specific halo with mass $M\ii$ at time $t\ii$
accreting at a given rate during any arbitrarily small time interval
$\Delta t$ around $t\ii$ is therefore simply the smooth extension
inward of the small piece of profile being built during that
interval.\footnote{The derivatives at any order of a smooth function,
which is known in some finite domain, are completely determined at any
point of that domain. Thus, by taking the Taylor series expansion of
the function in that point, one can extend it in a unique way outside
the domain.}
Unfortunately, the smooth extension of a small
piece of a function is hard to find in practice, so the mass profiles of
real individual halos can hardly be obtained in this way.

There is one case, however, in which such a smooth extension can
readily be achieved: that of halos with $M\ii$ at $t\ii$ {\it
accreting at the typical cosmological rate} $\ra(M(t),t)$ during any
arbitrarily small interval of time around $t\ii$. In this case, the
(unique) smooth extension we are looking for necessarily coincides
with the smooth function $M(t)$, the solution of the differential equation
\begin{equation}  
\frac{\dot M}{M(t)}=\ra(M(t),t)\, 
\label{mtrack}
\end{equation}
for the boundary condition $M(t\ii)=M\ii$, properly converted from $t$
to $r$ by means of equation (\ref{r}). Once the typical $M(r)$
profile is known, by differentiating it and taking into account
equations (\ref{r}) and (\ref{mtrack}), one is led to the {\it typical
density profile for halos with $M\ii$ at $t\ii$} proposed by M03,
\begin{equation}
\!\!\!\!\!\!\rho(r)=\frac{1}{4\pi r^2} \left(\frac{\dot M}{\dot R}\right)_{t(r)}
\!\!\!\!=\left[\delv(t)\bar \rho(t) \delta(t) \right]_{t(r)}
\label{rhot}
\end{equation}

\begin{equation}
\!\!\delta(t) =
\left[1-\frac{1}{\ra(M(t),t)}\frac{\der
\ln(\delv\bar \rho)}{\der t}\right]^{-1}\,.\label{delta} 
\end{equation}

\begin{figure}[htb]
	\centering
	\includegraphics[angle=0,width=0.5\textwidth]{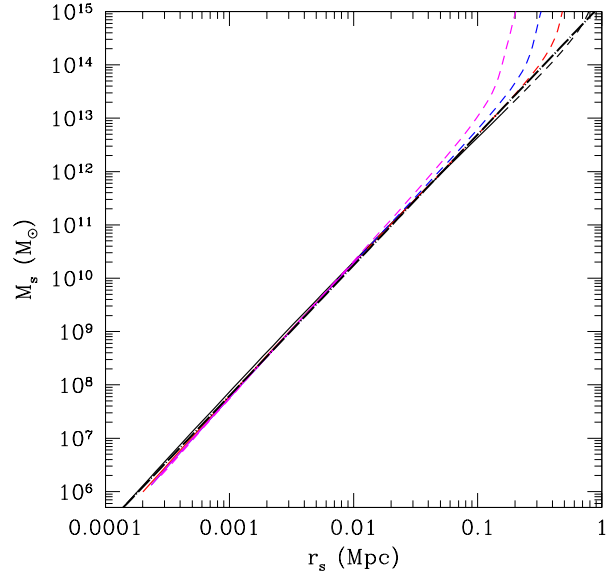}
	\vspace{-1 cm}
	\caption{Relations between the NFW $\rs$ and $M\s$ parameters obtained 
in the concordance model [sghown with solid lines for masses up to $10M_\ast(z)$
and dashed lines beyond it] from the fits of the theoretical
density profiles of halos with varying mass at the redshifts $0.0$,
1.03, 2.08, and 3.94. The dot-dashed line gives the analytical fit (see
text) to the mean of the solid lines.}
	\label{ms-rs}
\end{figure}

{}From equations (\ref{rhot}) and (\ref{delta}) we see that the shape
of this profile is ultimately set by the CDM power spectrum of density
perturbations in the cosmology considered through the merger rate
$\mr(M,t,\Delta)$ used to calculate the accretion rate $\ra(M,t)$ (see
eqs.~[\ref{mrate}] and [\ref{accrate}]). This dependence is, however, so
convoluted that the density profile (eq. [\ref{rhot}]) must be inferred
numerically. Only its central asymptotic behavior can be derived
analytically, as will be shown in the next section.

\section{Some consequences of the model}\label{dens}

{}From equation (\ref{accrate}) we see that the exact shape of the
density profile in equation (\ref{rhot}) depends on $\delm$. 
This parameter marks the effective transition between minor and
major mergers, and it can be determined from 
the empirical $M-c\/$ relation at some given redshift.

For each given $\delm$ value, the density profiles, down to $R/100$,
predicted at $z=0$ in the concordance model characterized by
$(\Omega\m,\Omega_\Lambda,h,\sigma_8)=(0.3,0.7,0.7,0.9)$ for halos
with different masses have been fitted to the NFW profile to find the
best-fit values of $c$. Then we searched for the
value of $\delm$ that minimizes the departure of the theoretical
$M-c\/$ relations from the empirical one drawn from
high-resolution simulations by \citet{Zetal03b}. As shown in Figure
\ref{fit}, $\delm=0.26$ gives an excellent fit over almost 4 decades
in mass ($8\times 10^{10} h^{-1}\modotm<M<4\times 10^{14}
h^{-1}\modotm$).

\begin{figure*}[htb]
	\centering
	\includegraphics[angle=0,width=0.8\textwidth]{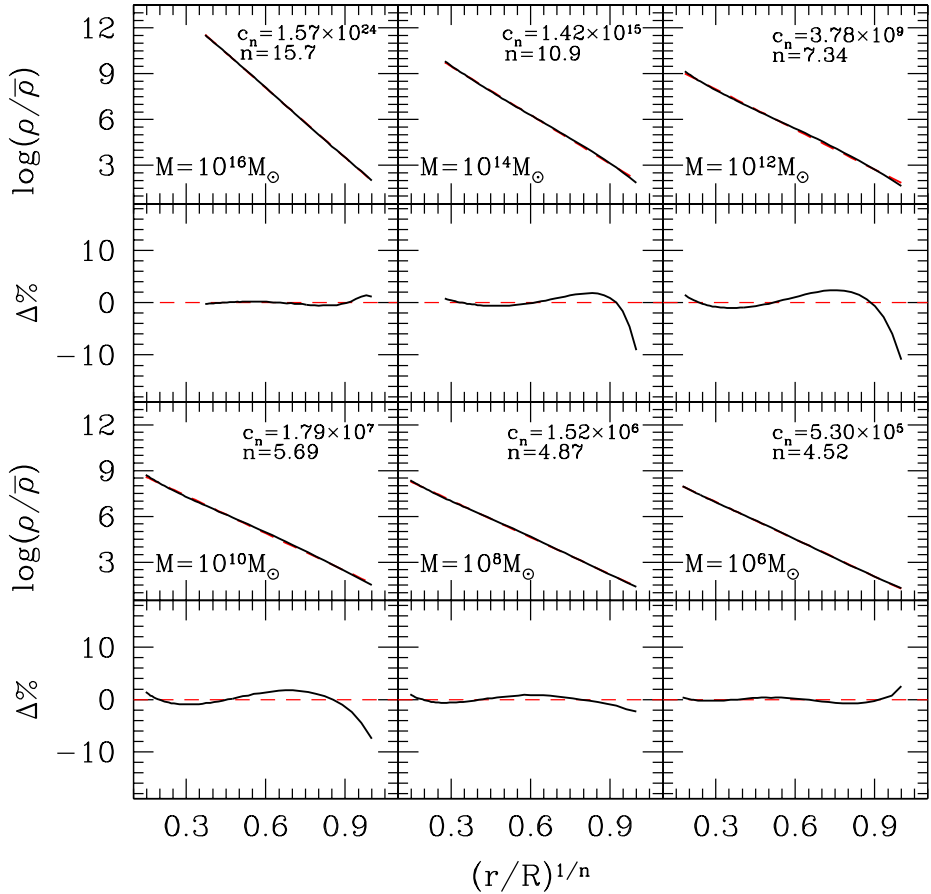}
	\vspace{-1.2cm}
	\caption{Same as Figure \ref{nfw}, but for the S\'ersic profile. All the
profiles are now drawn down to a 1 pc radius and cover a wider range of
halo masses: $10^{16} - 10^6\modotm$ (or, equivalently, from
$10^{3} M_{\star0}$ to $10^{-7} M_{\star0}$).}
	\label{sersic}
\end{figure*}

In Figure \ref{nfw}, we plot, down to a radius equal to the current
resolution radius of most numerical simulations, the density profiles
predicted in the concordance model for halo masses at $z=0$ ranging
from $10^{11}$ to $10^{16}\modotm$.  They are all well fitted to a NFW
profile, although there is a tendency for the theoretical profiles
for $M\ga 10^{14}\modotm$ to deviate from that shape and approach a
power law with logarithmic slope intermediate between the NFW
asymptotic values of $-1$ and $-3$. This tendency also makes $c$
increase very rapidly at large masses where the fit by the NFW law is
no longer acceptable. This causes the $M-c\/$ relation to deviate from
its regular trend at smaller values of $M$ (see Fig.~2). Both effects, already
reported in M03, were later observed in simulated halos
\citep{Zetal03b,Teal04}. This is a clear indication that the NFW
profile does not provide an optimal fit for very massive structures.
Of course, above 10$^{14}\modotm$ halos are hardly in virial
equilibrium, so such a deviation has essentially no practical effects.

As explained in \citet{SMS}, another interesting consequence of the
M03 model is that 
the $M(t)$ tracks traced by accreting halos (hence, growing inside
out) coincide with curves of constant $\rs$- and $M\s$- values, with
$M\s$ defined as the mass interior to $\rs$. Thus, the intersection of
those accretion tracks at any arbitrary redshift sets the relation
$M\s(\rs)$ between such a couple of parameters, implying that the
$M\s(\rs)$ relation satisfied by halos is {\it time-invariant}. In
Figure \ref{ms-rs}, we show how the different $M\s(\rs)$ curves
obtained by fitting the density profiles predicted at different
redshifts to a NFW law overlap. There is only some deviation at large
masses, where the density profiles are not correctly described by the
NFW profile. Such a time-invariant $M\s(\rs)$ relation is well fitted,
for $\rs$ in the range $10^{-4}~{\rm Mpc}< \rs< 1~{\rm Mpc}$, by
\begin{equation}
\frac{M\s}{10^{13}\modotm}=141\,\left(\frac{\rs}{\rm Mpc}\right)^{2.45}\,.
\label{rel0}
\end{equation}
Substituting equation  (\ref{rel0}) into the relation
\begin{equation}
\frac{M\s}{M}=\frac{\ln 2-0.5}{g(c)}\,
\label{ms}
\end{equation}
which holds for NFW profiles, one is led to
\begin{equation}
\!\!\!\!\!\!\left[\frac{4\pi\,\delv(t)\,\bar\rho(t)}{2.37\times 10^6\,\bar
\rho_0}\right]^{0.82}\!\left[\frac{M}{10^{13}\,M_\odot}\right]^{0.18}\!\!\!=
\frac{g(c)}{c^{2.45}}\,,
\label{rel}
\end{equation}
where $g(c)$ stands for $\ln(1+c)-c/(1+c)$ and $\bar \rho_0$ is the
current mean cosmic density. Equation (\ref{rel}) is an implicit
equation for the concentration of halos with any given mass and
redshift.

What about the central behavior of the predicted density profile?
According to equation (\ref{r}), small radii correspond to small
cosmic times. In this asymptotic regime, all Friedman cosmologies
approach the Einstein$-$de Sitter model in which $\delv(t)$ is constant
and $\bar\rho(t)$ is (in the matter-dominated era when halos form)
proportional to $t^{-2}$. If the  power spectrum of density
perturbations were of the power-law form $P(k)\propto k^j$, with the index $j$
satisfying $1>j> -3$ to guarantee hierarchical clustering, then the
universe would  be self-similar. The mass accretion in equation 
(\ref{mtrack}) would take the asymptotic form:
$M(t)\propto t^{\frac{2}{j+3}}$ (see the Appendix). The fact that both
$\delv(t)\bar\rho(t)$ and $M(t)$ would then be power laws has two
consequences. First, the time dependence of their
respective logarithmic derivatives on the right-hand side of equation
(\ref{delta}) cancel, which implies that $\rho(t)$ is proportional to
$\delv(t)\bar\rho(t)$, and hence to $\bar \rho(t)$. Second, the
virial radius given by equation (\ref{r}) is also a power law,
\begin{math}
R(t)\propto [M(t)/\bar\rho(t)]^{1/3}\propto t^{2(j+4)/[3(j+3)]}.
\end{math}
{}From this we get $t(r)$, and thereby one finds
\begin{equation}
\rho(r)\propto r^{-\frac{3(j+3)}{j+4}}\,.
\label{asrho}
\end{equation}
This central behavior, fully in agreement with the numerical profiles
obtained from power-law power spectra, is particularly robust, as it
does not depend on $\delm$. Note that it coincides with the solution
derived in self-similar cosmologies by \citet{HS85} assuming spherical
collapse. It is worth noting, however, that in that derivation, such an
asymptotic behavior is restricted to $j>-1$ so as to warrant the required
adiabatic invariance \citep{FG84}, while, in the present derivation,
there is no such a restriction as the central density profile is not
assumed to be built by spherical infall, but results instead from smooth
adaptation to the boundary condition imposed by current accretion.

The CDM power spectrum is, of course, not a power law. However, in the
limit of small masses involved in that asymptotic regime, it tends to
a power law of index $j=-3$. Thus, according to equation
(\ref{asrho}), we expect a vanishing central logarithmic slope of
$\rho(r)$ for the standard CDM case. This is confirmed by the
numerical profiles obtained in this case: as one goes deeper and
deeper into the halo center, they become increasingly shallower.

What is still more remarkable is that down to a radius as small as 1 pc, the
density profiles appear to be well fitted by the 3D S\'ersic or Einasto law
\begin{equation}
\rho(r)=\rho_0\exp\left[-\left(\frac{r}{r\n}\right)^{1/n}\right]\,,
\label{new}
\end{equation}
over at least 10 decades in halo mass (see Fig.~\ref{sersic}), from
$10^{6}\modotm$ to $10^{16}\modotm$.  We remind the reader that for power-law
spectra, equation (\ref{rhot}) leads to density profiles with central
cusps, so the S\'ersic shape is not a general consequence of the M03
model, but it is specific to the standard CDM power spectrum.  In
fact, from the reasoning above we see that what causes the zero
central logarithmic slope in the CDM case is the fact that the
logarithmic accretion rate $\der \ln M(t)/\der \ln t=t\ra(M(t),t)$
diverges in the limit of small values of $t$.  This is in contrast to the
general power-law case, where the accretion rate remains finite.

\begin{figure}[htb]
	\centering
	\includegraphics[angle=0,width=0.5\textwidth]{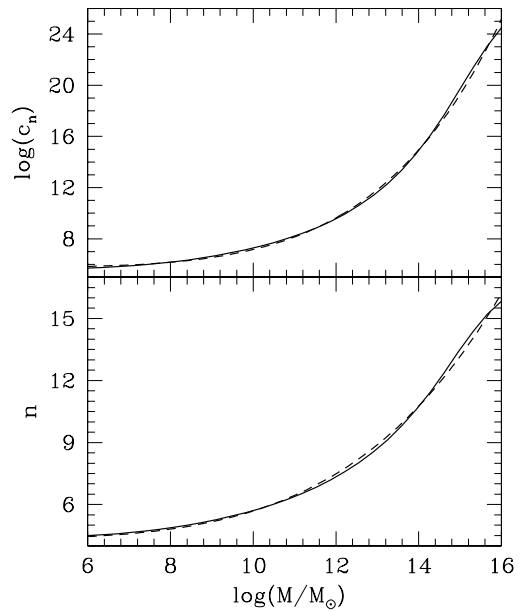}
	\caption{Mass dependence at $z=0$ of the S\'ersic law parameters $n$ and $c\n$
fitting the predicted density profiles of halos in the
concordance model ({\it solid lines}) and their fits by the simple
expressions given in the text ({\it dashed lines}).}
	\label{m-cn}
\end{figure}

Similar to the characteristic density $\rho\cc$ in the NFW profile
(eq.~[\ref{NFW}]), the central density $\rho_0$ entering the 3D
S\'ersic law (eq.~[\ref{new}]) can be written in terms of the mass $M$
and the values of the two (instead of one) remaining parameters, $n$
and either $r\n$ or $c\n\equiv R/r\n$:
\begin{equation}
\rho_0=\frac{M}{4\pi n r\n^3\Gamma(3n)}\,P^{-1}(3n,c\n^{1/n})\,,
\label{norm}
\end{equation}
where $\Gamma$ is the usual gamma function and
\begin{equation}
P(a,x)=\frac{1}{\Gamma(a)} \int_0^x \der t\; {\rm e}^{-t}\;t^{a-1}
\label{P}
\end{equation}
is the incomplete or regularized one. At $z=0$, the two free
parameters $n$ and $c\n$ depend on $M$ according to the relations
plotted in Figure \ref{m-cn}, which are well approximated by
\begin{equation}
n(M)=4.32+7.5\times 10^{-7}\left[\ln\left(\frac{M}{\modotm}\right)\right]^{4.6}\,,
\label{nm}
\end{equation}
\begin{equation}
\ln[c\n(M)]=13.3+
7.5\times 10^{-8}\left[\ln\left(\frac{M}{\modotm}\right)\right]^{5.6}\,,
\label{cnm}
\end{equation}
leading to the following combined relation
\begin{equation}
c\n(M)=5.84\times 10^5\left(\frac{M}{\modotm}\right)
^{\frac{n-4.32}{10}}\,.
\label{mild}
\end{equation}
These expressions can be used to infer the typical values of the
S\'ersic parameters for present-day halos with any  mass. It
is worth mentioning that the predicted values of $n$ are of the same
order of magnitude as the ones obtained by \citet{M05} from
simulated halos with masses ranging from dwarf galaxies to galaxy
clusters (the values of $c\n$ were not presented in that work).

\begin{figure}[htb]
	\centering
	\includegraphics[angle=0,width=0.5\textwidth]{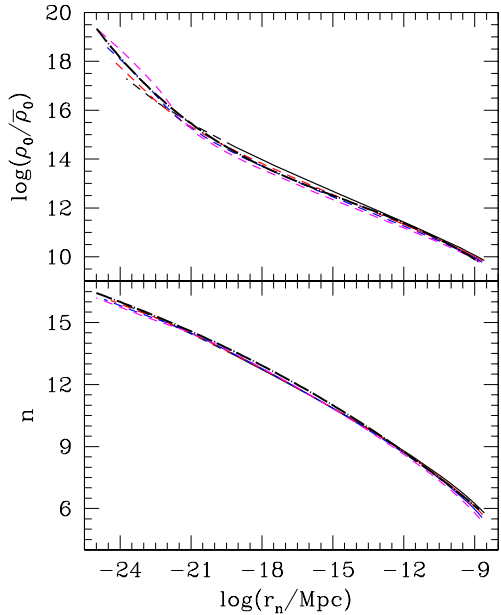}
	\caption{Same as Figure \ref{ms-rs} but for the $\rho_0(r\n)$
and $n(r\n)$ relations among the best-fit S\'ersic parameters. The
respective analytical fits (see text) to the mean curve are shown in
dot-dashed lines. Here $\bar\rho_0$ is the current mean cosmic density.}
	\label{sersic2}
\end{figure}

To obtain more general values of these parameters for halos of any
mass and redshift, we can proceed as in the NFW case
above. For reasons identical to the ones leading to the time-invariant
relation $M\s(\rs)$, the relations $\rho_0(r\n)$ and $n(r\n)$ must be
time-invariant. In Figure \ref{sersic2}, we see how the corresponding
curves obtained from the fit to the S\'ersic profile of the same density
profiles as used in Figure \ref{ms-rs} overlap, indeed, even better
than the $M\s(\rs)$ curves do, since there are no large deviations at large
masses. These invariant relations are well
fitted, for $x\equiv\log(r\n/{\rm Mpc})$ in the range $-25< x < -9$, by
\begin{eqnarray}
\rho_0(r\n)=&&\bar\rho_0\exp\left(A\right)  \, ,  \label{r1} \\
A =&& -2.48-4.73x \nonumber \\
&&-0.270x^2-6.24 
\times 10^{-3}x^3  \, , \nonumber
\end{eqnarray}
\begin{eqnarray}
n(r\n)=&&-4.43-1.43x \label{r2} \\
&&-0.0313x^2-2.98\times 10^{-4}x^3\,. \nonumber
\end{eqnarray}
Replacing these expressions into equation (\ref{norm}), we can solve
for $r\n$ and then use the equation (\ref{r2}) to find the value of
$n$.

This provides a very concrete prediction that can be tested with
numerical simulations. One can take the very strong correlation shown
in Figure \ref{sersic2}, which allows one to fit the density profile
of any dark matter structure with only two free parameters: e.g. $n$
and $\rho_0$. With this value of $n$ (purely from the {\it shape} of
the density profile), one now gets a value for the mass (from
Fig.~\ref{m-cn}).  This mass can trivially be compared to the true
virial mass (which is naturally known in the simulation), and hence
one can confirm or reject the prediction of the accretion-driven model.

\section{Summary}\label{sum}

We have shown how the basic properties of standard CDM can justify
the M03 model, which was previously shown to be in overall agreement with the
results of numerical simulations. In this model, the density profile
of relaxed halos permanently adapts to the profile currently building
up through accretion and does not depend on their past aggregation
history. As a consequence, the typical density profile of halos of a
given mass at a given epoch is set by their time-evolving
cosmology-dependent typical accretion rates.

Although halos have been assumed to be spherically symmetric
throughout the present paper, this is not crucial for the M03
model. As will be shown in a following paper (Gonz\'alez-Casado et al.~2007),
the results presented here also hold for more realistic triaxial
rotating halos. Furthermore, an approach similar to the one followed
here allows one to explain not only their mass distribution, but also
others of their structural and kinematic properties, such as the radial
dependence of angular momentum.

According to the M03 model, the central asymptotic behavior of the
halo density profile depends, through the typical accretion rate, on
the power spectrum of density perturbations. The prediction made in
the case of power-law spectra should be possible to check by means of
numerical simulations, provided one concentrates on massive halos, as
these reach the asymptotic regime at larger radii. In the case of the
standard CDM power spectrum, the model predicts a vanishing central
logarithmic slope. The way this asymptotic behavior is reached is
surprisingly simple: down to a radius as small as 1 pc, the density
profile is well fitted by the 3D S\'ersic or Einasto profile over at
least 10 decades in halo mass.

Another consequence of the M03 model with useful practical
applications is the existence of time-invariant relations among the
NFW or 3D S\'ersic law parameters ($M\s$ and $\rs$ in the former case
and $\rho_0$, $r\n$, and $n$ in the latter) fitting the halo density
profiles. A code is publicly available\footnote{\tt
http://www.am.ub.es/cosmo/gravitation.html} that computes such
invariant relations for any desired standard CDM cosmology.

Some of these consequences can be readily tested by numerical
simulations or by (X-ray or strong-lensing) observations, which should
allow one to confirm or reject the prediction for the central behavior
of the density profile of halos that is made by the M03 model.

\acknowledgments This work was supported by the Spanish DGES grant
AYA2006-15492-C03-03. We thank Donghai Zhao and co-workers for kindly
providing their data.

\appendix

\section{The accretion track asymptotic behavior}
\label{App1}

The instantaneous rate of mergers undergone by halos with mass $M$ at
time $t$ yielding a relative mass increase of $\Delta$, is \citep{LC93}
\begin{equation}
\mr(M,t,\Delta)=\frac{\sqrt{2/\pi}\,M}{\sigma^2(M')}\left|\frac{\der\delta\cc}
{\der t}\frac{\der\sigma(M')}{\der M'}\right|
\left[1-\frac{\sigma^2(M')}{\sigma^2(M)}\right]^{-3/2
}\exp\left\{-\frac{\delta\cc^2(t)}{2\sigma^2(M')}
\left[1-\frac{\sigma^2(M')}{\sigma^2(M)}\right]\right\},
\label{mrate} 
\end{equation}
where $M'\equiv M(1+\Delta)$ is the final mass, $\delta\cc(t)$ is the
linearly extrapolated, cosmology-dependent, critical density contrast
of density fluctuations at $t$ for collapse at the present time $t_0$,
and $\sigma(M)\equiv \sigma(M,t_0)$ is the rms density contrast of
fluctuations on scale $M$ at $t_0$, related to the zeroth-order
spectral moment.

In a self-similar universe with spectral index $j$, one has
\begin{equation}
\delta\cc(t)=\delta\cz\left(\frac{t}{t_0}\right)^{-2/3}\,,\quad
\sigma(M)=\sigma(M_{\star0})\left(\frac{M}{M_{\star0}}\right)^{-\frac{j+3}{6}}
\end{equation}
where $\delta\cz$ and $M_{\star0}$ are the current values of
$\delta\cc(t)$ and $M_\star(t)$, respectively, $M_\star(t)$ being the
critical mass for collapse at $t$, the solution of the implicit equation
$\sigma(M_{\star},t)=\delta\cz$. Therefore, equation (\ref{mrate})
takes the form
\begin{equation}
\mr(M,t,\Delta)= \sqrt{\frac{2}{\pi}}\,\frac{j+3}{9t}\,
\frac{\nu(M,t)\,(1+\Delta)^{2\frac{j+3}{3}-1}}
{\left[(1+\Delta)^{\frac{j+3}{3}}-1\right]^{3/2}}
\exp\left\{- \frac{\nu^2(M,t)}{2}
\left[(1+\Delta)^{\frac{j+3}{3}}-1\right]\right\},
\label{asymr}
\end{equation}
\begin{equation}
\nu(M,t)=\left(\frac{t}{t_0}\right)^{-2/3}\left(\frac{M}{M_{\star0}}\right)
^{\frac{j+3}{6}}=\left[\frac{M}{M_{\star}(t)}\right]
^{\frac{j+3}{6}}\,. 
\label{nu}
\end{equation}
Substituting equation (\ref{asymr}) into equation (\ref{accrate}) and
changing $\Delta$ by $x=\nu^2[(1+\Delta)^{(j+3)/3}-1]$, we
obtain the following expression for the accretion rate
\begin{equation}
\ra(M,t)=\frac{3A\nu^2}{(j+3)t}\,\int_0^{\xm(M,t,\delm)}
\der x\,\left[\left(1+\nu^{-2}x\right)^{\frac{3}{j+3}}-1\right]\nonumber
\left(1+\nu^{-2}x\right)\,
\frac{\exp\left(-x/2\right)}{x^{\frac{3}{2}}}\,,
\label{asrate}
\end{equation}
where, for simplicity, we have dropped the explicit dependence of
$\nu$ and used the notation $A=\sqrt{2/\pi}\,(j+3)/9$ and
$\xm(M,t,\delm)=\nu^2[(1+\delm)^{(j+3)/3}-1]$.

In the differential equation (eq. [\ref{mtrack}]), the variable $M$ in $\ra$
is replaced by the mass accretion track $M(t)$. As halos grow through
both accretion and major mergers, the accretion tracks $M(t)$ increase
with increasing time less rapidly than do $M_\star(t)$, tracing the
typical mass evolution of halos in any self-similar universe. Thus,
$\nu(t)\equiv \nu(M(t),t)$ is a decreasing function of $t$ (see
eq.~[\ref{nu}]) and, in the small-$t$ asymptotic regime, $\nu(t)^{-2}$
tends to zero. Taking the Taylor series expansion of
$[1+\nu^{-2}(t)x]^{3/(j+3)}$ inside the integral on the right-hand side of
equation (\ref{asrate}), at leading order in $\nu^{-1}(t)$, we obtain 
\begin{equation}
\ra(M(t),t)=\frac{9At^{-1}}{(j+3)^2}\int_0^{\xm(t)} \der x\,
\frac{\exp\left(-x/2\right)}{\sqrt x}
=\frac{9\sqrt{2\pi}At^{-1} }{(j+3)^2}
\;\erf{\left[\nu(t)\sqrt{(1+\delm)^{\frac{j+3}{3}}-1}\right]},
\label{asra}
\end{equation}
where $\erf(x)$ is the error function. Given the value of constant $A$,
and given that, for $\nu(t)$ tending to infinity, the error function
approaches unity, we are led to
\begin{equation}
\ra(M(t),t)=\frac{2}{j+3}\,t^{-1}\,.
\end{equation}
Finally, integrating equation (\ref{mtrack}) for such an accretion
rate, we are led to the following asymptotic behavior for accretion
tracks:
\begin{equation}
M(t)\propto t^{\frac{2}{j+3}}\,.
\end{equation}
Note how it compares with the time dependence of the critical mass in
a self-similar universe: $M_\star(t)\propto t^{\frac{4}{j+3}}$.

\end{document}